\documentclass[fleqn,twoside]{article}
\usepackage{espcrc2}

\usepackage{graphicx}
\usepackage[figuresright]{rotating}

\newcommand{\AmS}{{\protect\the\textfont2
  A\kern-.1667em\lower.5ex\hbox{M}\kern-.125emS}}

\newcommand{\be}{\begin{equation}}
\newcommand{\bea}{\begin{eqnarray}}
\newcommand{\ee}{\end{equation}}
\newcommand{\eea}{\end{eqnarray}}
\newcommand{\bpi}{\begin{picture}}
\newcommand{\bce}{\begin{center}}
\newcommand{\epi}{\end{picture}}
\newcommand{\ece}{\end{center}}

\def\g{\widetilde{\Gamma}}

\def\chic#1{{\scriptscriptstyle #1}}

\title{Scrutinizing the Green's functions of QCD: \\ Lattice meets Schwinger-Dyson}

\author{Joannis Papavassiliou \address{Department of Theoretical Physics and IFIC, \\
University of Valencia-CSIC,
E-46100, Valencia, Spain.}}%
              
\begin{document}

\begin{abstract}

The  Green's functions of  QCD 
encode  important  information  about  the infrared  dynamics  of  the
theory.  The main non-perturbative tools  used to study them are their
own  equations  of motion,  known  as  Schwinger-Dyson equations,  and
large-volume lattice  simulations.  We have now reached  a point where
the interplay between these two methods can be most fruitful.  Indeed,
the  quality  of the  lattice  data  is  steadily improving,  while  a
recently   introduced  truncation   scheme  for   the  Schwinger-Dyson
equations makes their  predictions far more reliable.  In  this talk 
several of the  above points will be reviewed, with particular emphasis on
how to enforce the crucial requirement of gauge invariance at the 
level of the  Schwinger-Dyson equations, 
the  detailed mechanism of  dynamical  gluon mass  
generation and its implications for the ghost sector,  
the non-perturbative effective charge of QCD, and the 
indirect extraction of the Kugo-Ojima function from existing lattice data.  

\vspace{1pc}
\end{abstract}

\maketitle

\section{Introduction}

The  basic  building  blocks  of  QCD are  the  Green's  (correlation)
functions of  the fundamental physical degrees of  freedom, gluons and
quarks, and  of the unphysical  ghosts.  Even though it  is well-known
that  these quantities  are not  physical,  since they  depend on  the
gauge-fixing scheme and the parameters used  to renormalize them, it is
widely  believed that reliable  information on  their non-perturbative
structure  is  essential  for  unraveling  the  infrared  dynamics  of
QCD. Indeed, in  addition  to  their  relevance  for
phenomenology,  the   QCD  Green's  function   encode  information  on
confinement,  albeit  in  a  rather  subtle  way.

The two  basic non-perturbative tools that permit the exploration of 
the infrared domain of QCD are
({\bf  i})  the  lattice,  where  space-time is  discretized  and  the
quantities of  interest are evaluated numerically, and  ({\bf ii}) the
infinite set of  integral equations governing the dynamics  of the QCD
Green's     functions,    known    as     Schwinger-Dyson    equations
(SDE). Given  that   both the lattice and the SDE aspire  to  describe
essentially the same physics, it  is important 
to advance their complementarity and 
strengthen their mutual interplay.
In fact, it  would seem  that  we have  reached  a point in time where  the
meaningful and systematic 
comparison between lattice and SDE results constitutes a  tangible  reality.
Indeed, the quality of lattice data is steadily improving, 
while,  due to several recent developments~\cite{Aguilar:2006gr},
we have at our disposal, for the first time,  
a manifestly  gauge invariant truncation scheme for the SD series of QCD.

It it  generally accepted by now  that the lattice  yields in the 
Landau  gauge  (LG)
an infrared finite  gluon  {\it propagator} 
and an infrared finite (non-enhanced) ghost {\it dressing function}.
This  rather characteristic  behavior has been  firmly established  recently using
large-volume lattices, for pure  Yang-Mills (no quarks included), for
both $SU(2)$~\cite{Cucchieri:2007md} and $SU(3)$~\cite{Bogolubsky:2007ud}.  
In this talk we will review the SD part of this story, 
within the
truncation scheme based on 
the pinch 
technique (PT)~\cite{Cornwall:1982zr,Cornwall:1989gv,Nair:2005iw} and 
its connection with the background field method (BFM)~\cite{Abbott:1980hw}. 
We discuss the plethora of conceptual issues and the wide 
range of possibilities that emerge 
when the dynamical gluon mass generation picture of QCD is adopted. 
Moreover, we clarify some basic but subtle issues related 
with the non-perturbative definition of the QCD effective charge.

\section{Problem(s) with the conventional SDEs}
The SDEs provide  a formal framework
for tackling physics  problems requiring a non-perturbative treatment.
Given that the SDEs constitute an infinite system of  coupled non-linear integral
equations for all Green's functions of the theory, 
their practical usefulness hinges crucially on one's ability to devise
a  self-consistent truncation  scheme  that would  select a  tractable
subset of these equations,  without compromising the physics one hopes
to describe.  
Devising  such  a  scheme,  however,  is  very
challenging, especially in the  context of non-Abelian gauge theories,
such as QCD.  
For the purposes of this presentation we will only focus on the 
aspect of the problem related with gauge-invariance 

In Abelian gauge theories the Green's functions satisfy naive 
Ward identities (WIs): the all-order 
generalization of a tree-level WI is obtained by simply replacing 
the Green's functions appearing in it by their all-order expressions. 
In general, this is not true in
non-Abelian gauge theories, where the WIs 
are modified non-trivially beyond tree-level, and are 
replaced by more complicated expressions known as       
Slavnov-Taylor identities  (STIs): 
in addition to the original Green's functions 
appearing at tree-level, they 
involve various composite {\it ghost operators}.

To appreciate why the WIs are instrumental for the  
consistent truncation of the SDEs, while the STIs complicate it, 
let us first consider how nicely things work in 
an Abelian case, namely {\it scalar} QED (photon), and then turn to the 
complications encountered in QCD (gluon). 

The full photon (gluon) propagator, in a $R_{\xi}$-type of gauge, is given by 
\begin{equation}
\Delta_{\mu\nu}(q)= -i\left[P_{\mu\nu}(q)\Delta(q^2) + 
\xi \frac{q_{\mu}q_{\nu}}{q^4}\right],
\label{prop_cov}
\end{equation}
where $P_{\mu\nu}(q)= g_{\mu\nu} - {q_\mu q_\nu}/{q^2}$ 
is the transverse projector.
The scalar function $\Delta(q^2)$ is related to the 
all-order (photon) gluon self-energy 
$\Pi_{\mu\nu}(q)=P_{\mu\nu}(q)\Pi(q^2)$
through $\Delta^{-1}(q^2) = q^2 + i\Pi(q^2)$.
We also define the dimensionless 
vacuum-polarization ${\bf \Pi}(q^2)$, as $\Pi(q^2)= q^2 {\bf \Pi}(q^2)$. 

Note now a crucial point: 
local gauge-invariance (BRST in the case of the gluon) forces 
$\Pi_{\mu\nu}(q)$ (photon and gluon alike) to satisfy the fundamental 
transversality relation
\be
q^{\mu} \Pi_{\mu\nu}(q) = 0 \,,
\label{fundtrans}
\ee
both perturbatively (to all orders) and non-perturbatively.

\begin{figure}[!t]
\includegraphics[scale=0.35]{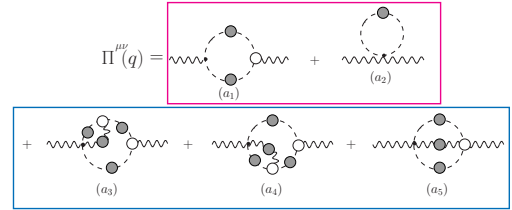}
\vspace{-1.0cm}
\caption{The SDE for the photon self-energy in scalar QED.}
\label{fig1}
\end{figure}

The SDE governing $\Pi_{\mu\nu}(q)$ in scalar QED is shown in Fig.\ref{fig1}.
The main question we want to address is the following: can one truncate the 
rhs of Fig.\ref{fig1}, i.e. eliminate some of the graphs, without compromising 
the transversality of $\Pi_{\mu\nu}(q)$ ? 
The answer is shown already in Fig.\ref{fig1}: the two blocks of graphs 
$[(a_1)+(a_2)]$ and $[(a_3)+(a_4)+(a_5)]$ are {\it individually transverse}, 
i.e.
\be  
q^{\mu}\sum_{i=1}^{2}(a_i)_{\mu\nu} = 0\,,\,\,\,\, q^{\mu}\sum_{i=3}^{5}(a_i)_{\mu\nu} = 0\,.
\ee
The reason for this special property 
are precisely the naive WIs satisfied by the full vertices 
appearing on the rhs of the SDE; for example, the first block is 
transverse simply because the {\it full} photon-scalar vertex $\Gamma_{\mu}$ [white blob in $(a_1)$]
satisfies the WI
\be
q^{\mu} \Gamma_{\mu} = e [{\mathcal D}^{-1}(k+q) -{\mathcal D}^{-1}(k)], 
\label{sgwi}
\ee
where ${\mathcal D}(q)$ is the {\it full} propagator of the charged scalar. 
A similar WI relating the four-vertex with a linear combination of $\Gamma_{\mu}$
forces the transversality of the second block in Fig.\ref{fig1}.
Thus, due to the simple WIs satisfied by the vertices appearing on the SDE for $\Pi_{\mu\nu}(q)$, 
one may omit the second block of graphs and still maintain the transversality of the 
answer intact, i.e. the approximate $\Pi_{\mu\nu}(q)$ obtained after this truncation 
satisfies (\ref{fundtrans}).

\begin{figure}[!t]
\includegraphics[scale=0.35]{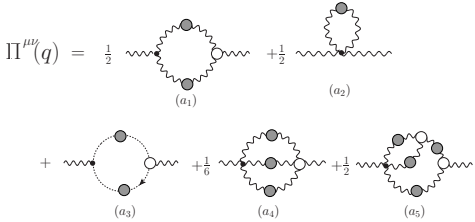}
\vspace{-1.0cm}
\caption{The conventional SDE for the gluon self-energy}
\label{fig2}
\end{figure}

\begin{figure}[!t]
\begin{center}
\includegraphics[scale=0.5]{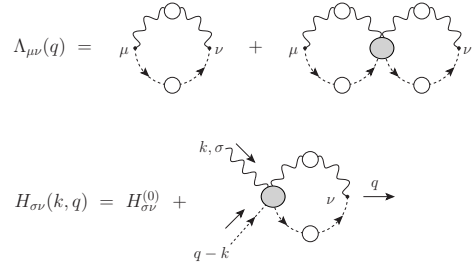}
\vspace{-1.0cm}
\caption{Diagrammatic representation of the functions $\Lambda$ and $H$.}
\label{fig3}
\end{center}
\end{figure}

Let us now turn to the {\it conventional} SDE for the gluon self-energy, 
in the $R_{\xi}$ gauges, given in Fig.\ref{fig2}.  
Clearly, by virtue of (\ref{fundtrans}), we must have 
\be
q^{\mu} \sum_{i=1}^{6}(a_i)_{\mu\nu} =0 . 
\label{abcde}
\ee
However, unlike the Abelian example, the diagrammatic verification 
of (\ref{abcde}), i.e. through contraction of the individual graphs by $q^{\mu}$, 
is very difficult, essentially due to the 
complicated STIs satisfied by the vertices involved. 
For example, the full three-gluon vertex $\Gamma_{\alpha\mu\nu}(q,k_1,k_2)$ satisfies 
the STI
\bea
q^\alpha \Gamma_{\alpha\mu\nu} &=&
F(q) \Delta^{-1}(k_2^2) P^{\gamma}_{\nu}(k_2) H_{\mu\gamma}(k_1,k_2)
\nonumber\\
&-& F(q)\Delta^{-1}(k_1^2) P^{\gamma}_{\mu}(k_1) H_{\nu \gamma}(k_2,k_1),
\label{sti3gv}
\eea
where $F(q) = q^2 D(q)$ is the ghost dressing function,  $D(q)$ is the 
ghost propagator, and  $H_{\mu\nu}$, defined in Fig.\ref{fig3},  
is related to the full gluon-ghost vertex by 
$q^\nu H_{\mu\nu}(k,q)=-i\Gamma_{\nu}(k,q)$; at tree-level, $H_{\mu\nu}^{(0)} = ig_{\mu\nu}$.
In addition, some of the pertinent STIs are either too complicated,
such as that of the conventional four-gluon vertex, or they 
cannot be cast in a particularly convenient  form, such as 
the  STI of the conventional gluon-ghost vertex.
The main practical consequence of all this 
is that one {\it cannot} truncate the rhs of Fig.\ref{fig2} in any obvious way 
without violating the transversality of the resulting $\Pi_{\mu\nu}(q)$.
For example, keeping only graphs $(a_1)$ and $(a_2)$ is not correct even perturbatively,
since the ghost loop is crucial for the transversality of  $\Pi_{\mu\nu}$
already at one-loop;
adding $(a_3)$ is still not sufficient for a SDE analysis, because
(beyond one-loop) \mbox{$q^{\mu}[(a_1)+(a_2) + (a_3)]_{\mu\nu} \neq 0$}.

\section{Truncating gauge-invariantly: \\The PT-BFM framework}
\begin{figure}[!t]
\includegraphics[scale=0.35]{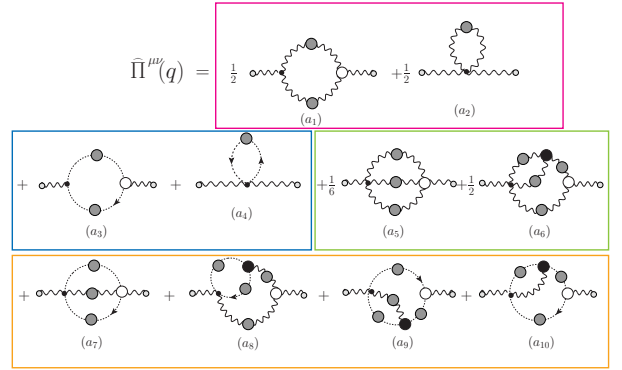}
\vspace{-1.5cm}
\caption{The SDE for the gluon self-energy in the PT-BFM framework.}
\label{fig4}
\end{figure}

Recently a new truncation scheme for the SDEs of QCD has been proposed, 
that respects gauge  invariance  at every level  of  the ``dressed-loop''
expansion.  This becomes  possible  due to  the drastic  modifications
implemented  to  the building  blocks  of  the  SD series,  {\it i.e.}   the
off-shell  Green's  functions themselves,  following 
the  general methodology of the PT~\cite{Cornwall:1982zr,Cornwall:1989gv,Nair:2005iw}.
The PT  is a well-defined  algorithm that exploits  systematically the
BRST symmetry in order to construct new Green's functions endowed with
very  special  properties.  In particular, the crucial property for 
our discussion is that they satisfy  Abelian WIs instead  of the usual STIs
The PT may be used to rearrange systematically the entire SD series~\cite{Aguilar:2006gr}. In the case of  
the gluon self-energy it 
gives rise {\it dynamically} to a new SDE, Fig.\ref{fig4}, similar to that of Fig.\ref{fig2}, 
but with two important differences:

{\bf (i)} What appears on the lhs is {\it not} 
the conventional self-energy $\Pi_{\mu\nu}$, but rather 
the PT-BFM self-energy, denoted by $\widehat{\Pi}_{\mu\nu}$.
Of course, $\widehat\Pi_{\mu\nu}$ is also transverse, i.e. 
$q^{\mu} \widehat{\Pi}_{\mu\nu}(q) = 0$. Quite interestingly, 
the two quantities are 
connected by a powerful formal identity~\cite{Grassi:1999tp}
stating that 
\be
\Delta(q^2) = 
\left[1+G(q^2)\right]^2 \widehat{\Delta}(q^2), 
\label{bqi2}
\ee
with $G(q^2)$ defined from $\Lambda_{\mu \nu}(q)$, Fig.\ref{fig3}, 
\bea
i\Lambda_{\mu \nu}(q) \!\!\!\! &=& \!\!\!\!\lambda 
\int_k H^{(0)}_{\mu\rho}
D(k+q)\Delta^{\rho\sigma}(k)\, H_{\sigma\nu}(k,q),
\nonumber \\
&=& \!\!\! g_{\mu\nu} G(q^2) + \frac{q_{\mu}q_{\nu}}{q^2} L(q^2),
\label{LDec}
\eea
where \mbox{$\lambda=g^2 C_{\rm {A}}$}, with $C_{\rm {A}}$ the Casimir eigenvalue of the adjoint representation
[$C_{\rm {A}}=N$ for $SU(N)$], 
and \mbox{$\int_{k}\equiv\mu^{2\varepsilon}(2\pi)^{-d}\int\!d^d k$}, 
with $d=4-\epsilon$ the dimension of space-time.


{\bf (ii)} The graphs appearing on the rhs  
contain the conventional self-energy $\Pi_{\mu\nu}$ as before, 
but are composed out of new vertices (Fig.\ref{fig4}).
These new vertices correspond 
precisely to the Feynman rules of the BFM~\cite{Abbott:1980hw}, {\it i.e.},
it is as if the external gluon had been converted 
dynamically into a background gluon. As a result,
the full vertices ${\g}_{\alpha\mu\nu}^{amn}(q,k_1,k_2)$, ${\g}_{\alpha}^{anm}(q,k_1,k_2)$,
${\g}_{\alpha\mu\nu\rho}^{amnr}(q,k_1,k_2,k_3)$, and ${\g}_{\alpha\mu}^{amnr}(q,k_1,k_2,k_3)$
appearing on the rhs of the SDE shown in Fig.\ref{fig4} satisfy the simple WIs 
\bea
q^{\alpha}{\g}_{\alpha\mu\nu}^{amn} &=&
gf^{amn}
\left[\Delta^{-1}_{\mu\nu}(k_1)
- \Delta^{-1}_{\mu\nu}(k_2)\right],
\nonumber\\
q^{\alpha}{\g}_{\alpha}^{anm} &=&  igf^{amn}
\left[D^{-1}(k_1)- D^{-1}(k_2)\right],
\nonumber\\ 
q^{\alpha}{\g}_{\alpha\mu\nu\rho}^{amnr} &=&
g f^{adr} {\Gamma}_{\nu\rho\mu}^{drm}(q+k_2,k_3,k_1) + {\rm cp},
\nonumber\\
q^{\alpha} {\g}_{\alpha\mu}^{amnr} &=&
g f^{aem} \Gamma_{\mu}^{enr}(q+k_1,k_2,k_3) + {\rm cp}. 
\label{fourWI}
\eea
where ``cp'' stands for ``cyclic permutations'' 
The implication of this for the truncation of the SDE are far-reaching. Indeed,
using these WIs, it is elementary to show that~\cite{Aguilar:2006gr} 
\bea
&& q^{\mu}\sum_{i=1}^{2}(a_i)_{\mu\nu} = 0\,,\,\,\,\, q^{\mu}\sum_{i=3}^{4}(a_i)_{\mu\nu} = 0\,,
\nonumber\\
&& q^{\mu}\sum_{i=5}^{6}(a_i)_{\mu\nu} = 0\,,\,\,\,\, q^{\mu}\sum_{i=7}^{10}(a_i)_{\mu\nu} = 0\,.
\label{tranbl}
\eea
Evidently, the SDE is composed of ``one-loop'' and ``two-loop'' dressed blocks that are 
{\it individually transverse}, exactly as happened 
in the Abelian case. In fact, the resulting pattern is even more spectacular: 
the gluon and ghost diagrams form separate transverse blocks!

\section{Gluon mass generation: the big picture}

Even though  the  gluon is
massless  at the  level  of the  fundamental  QCD Lagrangian, and  remains
massless to all order in perturbation theory, the non-perturbative QCD
dynamics  generate  an  effective,  momentum-dependent  mass,  without
affecting    the   local    $SU(3)_c$   invariance,    which   remains
intact~\cite{Cornwall:1982zr}.    

The gluon mass generation is a purely non-perturbative effect.
At the level of the SDEs
the  generation of such a  mass is associated with 
the existence of 
infrared finite solutions for the gluon propagator~\cite{Cornwall:1982zr,Aguilar:2008xm}, 
i.e. solutions with  $\Delta^{-1}(0) > 0$ .
Such solutions may  
be  fitted  by     ``massive''  propagators  of   the form 
$\Delta^{-1}(q^2)  =  q^2  +  m^2(q^2)$;
$m^2(q^2)$ is  not ``hard'', but depends non-trivially  on the momentum  transfer $q^2$.
In addition, one obtains   
the  non-perturbative  generalization  of  $\alpha(q^2)$, 
the  QCD  running  coupling (effective charge), in the form~\cite{Cornwall:1982zr}
\be
\alpha^{-1}(q^2)= b \ln\left(\frac{q^2+\,4\, m^2(q^2)}{\Lambda^2}\right)\,.
\ee 
The $m^2(q^2)$ in the argument of the logarithm 
tames  the   Landau pole, and $\alpha(q^2)$ freezes 
at a  finite value in the IR, namely  
\mbox{$\alpha^{-1}(0)= b \ln (4m^2(0)/\Lambda^2)$}.
Moreover, the gluon mass forces $F(q^2)$ to stay IR-finite (non-enhanced)~\cite{Aguilar:2008xm,Boucaud:2008ji}. 

{\it (i) The Schwinger mechanism}

In order to obtain  massive solutions {\it gauge-invariantly}, 
it  is necessary to invoke the well-known Schwinger mechanism~\cite{Schwinger:1962tn}:    
if, for some reason, ${\bf \Pi}(q^2)$ 
acquires a pole at zero momentum transfer, then the 
 vector meson becomes massive, even if the gauge symmetry 
forbids a mass at the level of the fundamental Lagrangian.
Indeed, it is clear that if the vacuum polarization ${\bf \Pi}(q^2)$ has a pole at  $q^2=0$ with positive 
residue $\mu^2$, i.e. ${\bf \Pi}(q^2) = \mu^2/q^2$, then (in Euclidean space)
$\Delta^{-1}(q^2) = q^2 + \mu^2$.
Thus, the vector meson 
becomes massive, $\Delta^{-1}(0) = \mu^2$, 
even though it is massless in the absence of interactions ($g=0$). 
There is {\it no} physical principle which would preclude ${\bf \Pi}(q^2)$ from 
acquiring such a pole.
In a {\it strongly-coupled} theory like QCD 
this may happen for purely dynamical reasons,
since strong binding may generate zero-mass bound-state excitations~\cite{Jackiw:1973tr}.
The latter  act  {\it  like}
dynamical Nambu-Goldstone bosons, in the sense that they are massless,
composite,  and {\it longitudinally   coupled};  but, at  the same  time, they
differ  from  Nambu-Goldstone  bosons   as  far  as  their  origin  is
concerned: they  do {\it not} originate from  the spontaneous breaking
of  any global symmetry~\cite{Cornwall:1982zr}.

\begin{figure}[!t]
\begin{center}
\includegraphics[scale=0.4]{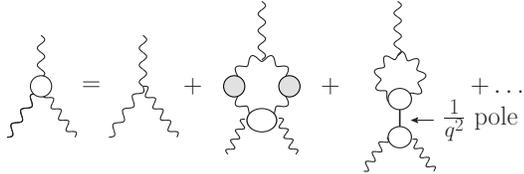}
\vspace{-1.0cm}
\caption{Vertex with  non-perturbative massless excitations triggering the Schwinger mechanism.}
\label{fig5}
\end{center}
\end{figure}

As we will see in a moment, the exact way how the Schwinger mechanism 
is integrated into the SDE   
is through the form of the three-gluon vertex. 
In particular,   
one assumes that the vertex contains {\it dynamical poles}
$\sim {1}/{q^2}$ [see Fig.\ref{fig5}], which will trigger the Schwinger mechanism when  
inserted into the SDE for the gluon self-energy
(other vertices, e.g, the four-gluon vertex, may have similar poles).

{\it (ii) Gluon mass and confinement} 

It has been occasionally argued that the concept of a massive gluon may be 
at odds with confinement, because a massive gauge field does not give rise 
to the long-range potential necessary for this latter phenomenon to occur.
This is, however, not true:
{\it the gluon mass does not obstruct confinement, it enables it!}

Of course, the exact way how this happens is very intricate, 
and is inextricably connected with the notion of a {\it quantum soliton}.
A quantum soliton is a localized finite-energy configuration of gauge potentials arising from
an effective action that summarizes quantum effects not present in the
classical action; in our case, the quantum effect is a gauge-invariant dynamical
gluon mass. Specifically, an effective low-energy field 
theory for describing the gluon mass   
is  the  gauged non-linear sigma model  known  as ``massive
gauge-invariant Yang-Mills''~\cite{Cornwall:1979hz}, with    
Lagrangian density  ${\cal L}_m$ 
\begin{equation}
{\cal L}_m= \frac{1}{2} G_{\mu\nu}^2 - 
m^2 {\rm Tr} \left[A_{\mu} - {g}^{-1} U(\theta)\partial_{\mu} U^{-1}(\theta) \right]^2,
\label{nlsm}
\end{equation}
where 
$A_{\mu}= \frac{1}{2i}\sum_{a} \lambda_a A^{a}_{\mu}$, the $\lambda_a$ are the SU(3) generators
(with  ${\rm Tr} \lambda_a  \lambda_b=2\delta_{ab}$), 
and the $N\times N$
unitary matrix $U(\theta) = \exp\left[i\frac{1}{2}\lambda_a\theta^{a}\right]$ 
describes the scalar fields $\theta_a$.  
${\cal L}_m$ is locally gauge-invariant under the combined gauge transformation 
\bea
A^{\prime}_{\mu} &=& V A_{\mu} V^{-1} - {g}^{-1} \left[\partial_{\mu}V \right]V^{-1}\,, 
\nonumber\\
U^{\,\prime} &=& U(\theta^{\,\prime}) = V U(\theta)\,,
\label{gtransfb}
\eea
for any group matrix $V= \exp\left[i\frac{1}{2}\lambda_a\omega^{a}(x)\right]$, where 
$\omega^{a}(x)$ are the group parameters. 
${\cal L}_m$ admits  vortex
solutions -- these are the aforementioned quantum solitons --
  with a  long-range pure  gauge term  in  their potentials,
which endows  them with a topological quantum  number corresponding to
the center  of the gauge group  [$Z_N$ for $SU(N)$], and  is, in turn,
responsible for quark  confinement and gluon screening. 
Specifically, center vortices of  thickness $\sim m^{-1}$,  where $m$ is
the induced mass of the gluon, form a condensate because their entropy
(per  unit  size) is  larger  than  their  action.  This  condensation
furnishes an  area law to  the fundamental representation  
Wilson loop, thus confining quarks~\cite{Cornwall:1979hz}.


\begin{figure}[tbp]
\begin{center}
\includegraphics[scale=0.5]{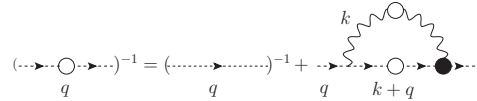}
\vspace{-1.0cm}
\caption{The SDE for the ghost propagator.}
\label{fig6}
\end{center}
\end{figure}

\begin{figure*}[tbp]
\begin{center}
\includegraphics[scale=2.0]{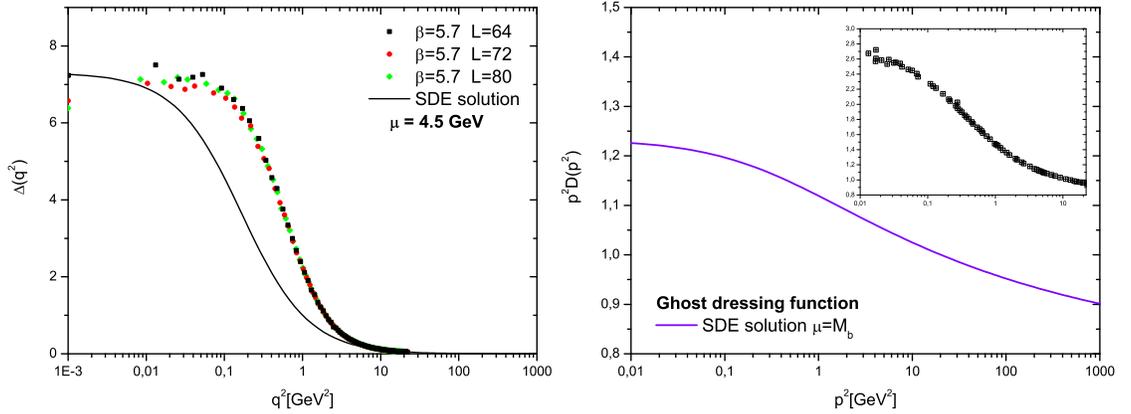}
\vspace{-1.5cm}
\caption{Comparison between SDE results of~\cite{Aguilar:2008xm} and the lattice results of~\cite{Bogolubsky:2007ud}.}
\label{fig7}
\end{center}
\end{figure*}

\section{Solutions of the SDEs and comparison with the lattice results}

We are now in a position to incorporate into the SDE truncation scheme 
based on the PT-BFM, described above, the concrete dynamics that will give rise to 
an infrared finite gluon propagator, signaling the 
dynamical generation of a gluon mass. To that end, we truncate (gauge-invariantly!) 
the SDE for the gluon propagator by keeping only the one-loop dressed contributions, i.e.
the first two blocks of graphs in Fig.\ref{fig4}. Then, we have~\cite{Aguilar:2008xm} 
\be
\Delta^{-1}(q^2){\rm P}_{\mu\nu}(q) = 
\frac{q^2 {\rm P}_{\mu\nu}(q) + i\,\sum_{i=1}^{4}(a_i)_{\mu\nu}}{[1+G(q^2)]^2}.
\label{sd1}
\ee

We next express ${\g}_{\mu\alpha\beta}(q,k_1,k_2)$ 
and ${\g}_{\mu}(q,k_1,k_2)$ [appearing in $(a_1)$ and $(a_3)$ of Fig.\ref{fig4}, respectively]
as a function of the gluon and ghost self-energy, respectively, in such a way as to 
automatically satisfy the first two WIs of (\ref{fourWI}); failure to do so 
would invariably compromise the transversality of the answer.  
The Ansatz we will use for ${\g}_{\alpha\mu\nu}(q,k_1,k_2)$ is 
\be
{\g}_{\mu\alpha\beta}= \Gamma_{\mu\alpha\beta}^{(0)} + i\frac{q_{\mu}}{q^2}
\left[\Pi_{\alpha\beta}(k_2)-\Pi_{\alpha\beta}(k_1)\right]\,,
\label{gluonv}
\ee
and a similar expression for ${\g}_{\mu}$. 
The essential feature
is the presence of massless 
pole terms, required  for triggering the Schwinger mechanism.
The resulting final expression for the SDE of (\ref{sd1})
is too lengthy to report here~\cite{Aguilar:2008xm}. 
The main point is that, as desired, $\Delta^{-1}(0) > 0$, i.e. 
the gluon self-energy is infrared finite.

For the conventional ghost-gluon vertex ${\Gamma}_{\mu}$,
appearing in the ghost SDE of Fig.\ref{fig6}, we use its
tree-level expression, {\it i.e.}, $\Gamma_{\mu} = -p_{\mu}$;
this is perfectly legitimate, since in the PT-BFM formalism 
the two ghost vertices,  ${\g}_{\mu}$ and ${\Gamma}_{\mu}$, are different.
Finally, for $H_{\mu\nu}$ we use its tree-level value, $ig_{\mu\nu}$. 

Then, we obtain (Euclidean space)~\cite{Aguilar:2008xm,Aguilar:2009nf}
\bea
F^{-1}(q^2) \!\!\! &=& \!\!\! 1 - \lambda \int_k\,\left[1-\frac{(q \cdot k)^2}{q^2 k^2}
\right]\Delta(k)\,D(q+k),
\nonumber\\
G(q^2) \!\!\! &=& \!\!\!\! - \frac{\lambda}{3}\int_k\,\left[
2+\frac{(q\cdot k)^2}{k^2q^2}\right]\Delta(k)D(k+q),
\nonumber\\
L(q^2) \!\!\! &=& \!\!\!\!
-\frac{\lambda}{3} \int_k \left[1 - 4\frac{(q \cdot k)^2}{k^2 q^2}\right]\Delta (k)  D(k+q). 
\nonumber\\
&&{}
\label{sd3}
\eea

In fact, there exists a powerful formal identity relating $F(q^2)$, $G(q^2)$, and $L(q^2)$, namely~\cite{Grassi:2004yq}  
\be
F^{-1}(q^2) = 1+G(q^2)+L(q^2)\,.
\label{ids}
\ee
In addition to its formal derivation~\cite{Grassi:2004yq}, 
the above relation has been recently obtained at the level of the SDEs defining these 
three quantities~\cite{Aguilar:2009nf}. 
Adding by parts Eqs.(\ref{sd3}) we can verify 
that Eq.(\ref{ids}) is indeed satisfied under the approximations employed. 

Eq.(\ref{ids}) merits further analysis.
The origin of this equation  
is the BRST symmetry of the theory~\cite{Grassi:2004yq}; in that sense,  Eq.(\ref{ids}) has 
the same nature 
as the STIs. Therefore, just as happens with 
the latter, (\ref{ids}) must not get deformed after renormalization~\cite{Aguilar:2009nf}.
Thus, denoting by $Z_u$  the 
renormalization constant relating 
the bare and renormalized functions, $\Lambda_0^{\mu\nu}$ and $\Lambda^{\mu\nu}$, [see (\ref{LDec})]
through
\be
Z_u [g^{\mu\nu} + \Lambda_0^{\mu\nu}(q)] = 
g^{\mu\nu} +\Lambda^{\mu\nu}(q), 
\label{Lamrel}
\ee
the requirement of non-deformation imposes the crucial condition $Z_u = Z_{c}$~\cite{Aguilar:2009nf}. Thus,  
\mbox{$Z_c [1+G_0(q^2,\Lambda^2)]=1+G(q^2,\mu^2)$} and \mbox{$Z_c L_0(q^2,\Lambda^2) = L(q^2,\mu^2)$}. 

The solutions obtained form the above system of SDEs 
is shown in Fig.\ref{fig7}, where it is compared with the 
corresponding lattice data [$L(q^2)$ is numerically suppressed and is not shown here].
Note that while there is good qualitative agreement with the lattice, 
there is a significant discrepancy (a factor of 2) in the intermediate 
region of momenta. This of course may not come as a surprise, given that  
the ``two-loop'' dressed part of the 
SDE for $\Delta$  
has been omitted [the last two blocks in Fig.\ref{fig4}].
Even though this omission has not 
introduced artifacts (since it was done gauge-invariantly), 
the terms left out
are expected to modify precisely the intermediate region, given that 
both the IR and UV limits of the solutions are already 
captured by the terms considered here.   

\begin{figure*}[!t]
\begin{center}
\includegraphics[scale=2.0]{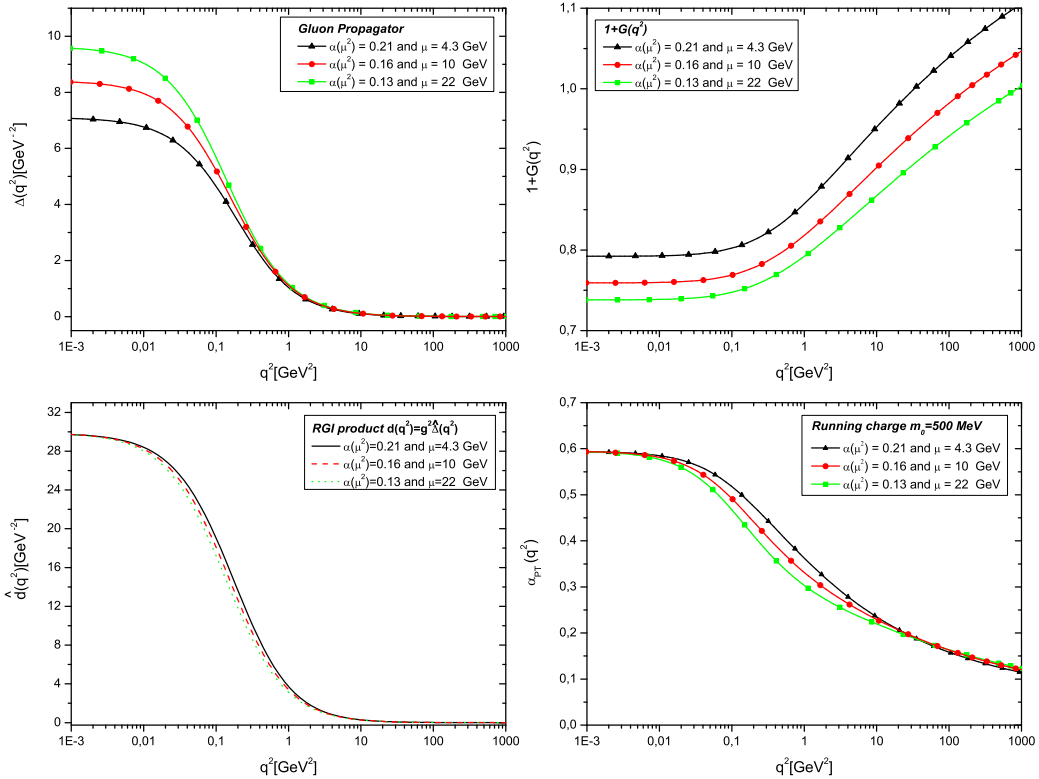}
\vspace{-2.5cm}
\caption{{\it Top left}: $\Delta(q^2)$ in the LG;
{\it Top right}: the $1+G(q^2)$;
{\it Bottom left}: the RG-invariant dimensionful $\widehat{d}(q^2)$ ;
{\it Bottom right}: The ${\alpha}_{\chic{\mathrm{PT}}}(q^2)$ 
obtained from $\widehat{d}(q^2)$, using (\ref{dwg}) and (\ref{rmass}), with \mbox{$m_0=500 \,\mbox{MeV}$}.}
\label{fig8}
\end{center}
\end{figure*}

\section{The effective charge of QCD}

Due to the Abelian WIs satisfied by the PT-BFM Green's functions, the 
 new $\widehat\Delta^{-1}(q^2)$ absorbs all the  
renormalization-group (RG) logarithms, 
exactly as happens in QED with the photon self-energy~\cite{Cornwall:1982zr,Cornwall:1989gv,Nair:2005iw}.
As a result, the renormalization 
constants of the gauge-coupling and of the PT gluon self-energy, 
defined as 
\bea
g(\mu^2) &=&Z_g^{-1}(\mu^2) g_0 ,\nonumber \\
\widehat\Delta(q^2,\mu^2) & = & \widehat{Z}^{-1}_A(\mu^2)\widehat{\Delta}_0(q^2), 
\label{conrendef}
\eea
where the ``0'' subscript indicates bare quantities, satisfy the 
QED-like relation ${Z}_{g} = {\widehat Z}^{-1/2}_{A}$.
Thus, regardless of  the renormalization prescription chosen, 
the product 
\be
{\widehat d}_0(q^2) = g^2_0 \widehat\Delta_0(q^2) = g^2(\mu^2) \widehat\Delta(q^2,\mu^2) = {\widehat d}(q^2) 
\label{ptrgi}
\ee
retains the same form before and after renormalization, {\it i.e.}, it 
forms a RG-invariant ($\mu$-independent) quantity~\cite{Cornwall:1982zr}.
Note that, by virtue of (\ref{bqi2}), we have that 
\be
{\widehat d}(q^2) = g^2(\mu^2)\frac{\Delta(q^2,\mu^2)}{[1+G(q^2,\mu^2)]^2}. 
\label{dwg}
\ee 
To see how beautifully the $\mu$-independence of ${\widehat d}(q^2)$  
is captured by the solutions of the SDEs we study, note in Fig.\ref{fig8}
the explicit $\mu$-dependence of the individual ingredients entering into its
definition. 
However, when they are put together according to the rhs of (\ref{dwg}), the resulting 
${\widehat d}(q^2)$ is practically $\mu$-independent!  

The next step is to extract out of ${\widehat d}(q^2)$
a {\it dimensionless} quantity, that would correspond to the QCD {\it effective charge}.
Perturbatively, i.e. for asymptotically large momenta, 
it is clear that the mass scale is saturated simply by $q^2$, the bare gluon propagator,  
and the effective charge is defined by pulling a $1/q^{2}$ out of the corresponding  
RG-invariant quantity, according to 
\be
{\widehat d}(q^2) = \frac{\overline{g}^2(q^2)}{q^2},
\label{ddef1}
\ee
where $\overline{g}^2(q^2)$ is the RG-invariant effective charge of QCD; at one-loop
\be
\overline{g}^2(q^2) = \frac{g^2}{1+  b g^2\ln\left(q^2/\mu^2\right)}
= \frac{1}{b\ln\left(q^2/\Lambda^2_\chic{\mathrm{QCD}}\right)}.
\label{effch}
\ee
where $\Lambda_\chic{\mathrm{QCD}}$ denotes an RG-invariant mass scale of a few hundred ${\rm MeV}$.

Of course, given that  
in the IR the gluon propagator becomes effectively massive, 
{\it particular care is needed in deciding exactly what 
combination of mass-scales ought to be pulled out}.
It would certainly be unwise, for example, to continue defining the effective charge 
by  forcing out just a factor of $1/q^2$; 
such a procedure would furnish (trivially) a completely unphysical coupling, 
vanishing in the deep infrared, where QCD is supposed to be strongly coupled !
The correct procedure, instead, is to factor out 
a ``massive'' propagator, of the form  \mbox{$[q^2 + m^2(q^2)]^{-1}$},  
i.e. one must set~\cite{Cornwall:1982zr,Aguilar:2009nf}
\be
\widehat{d}(q^2) = \frac{\overline{g}^2(q^2)}{q^2 + m^2(q^2)} \,,
\label{ddef}
\ee
where $m^2(q^2)$ is a momentum-dependent gluon mass.
Clearly, for \mbox{$q^2\gg m^2(q^2)$} 
the expression on the rhs of (\ref{ddef}) goes over to that of (\ref{ddef1}).
The effective charge, ${\alpha}_{\chic{\mathrm{PT}}}(q^2) =  \overline{g}^2(q^2)/4 \pi$,  
obtained from the $\widehat{d}(q^2)$ after using (\ref{ddef}),
is also shown in Fig.\ref{fig8}.
We have assumed ``power-law'' running of the mass, 
\be
m^2(q^2) = \frac{m_0^4}{q^2+m_0^2} \,,
\label{rmass}
\ee
consistent with various independent studies~\cite{Lavelle:1991ve}.

Let us finally comment on some important conceptual issues.
As already mentioned, the 
``freezing'' of the coupling is a direct consequence of the 
appearance of a mass in the RG logarithm. 
This fundamental property of the strong coupling may be reformulated 
in terms of what  in the language of the {\it effective field theories} is referred to as ``decoupling'':
at energies sufficiently  inferior to their masses, the particles appearing in the loops  
(in this case  the gauge bosons) 
cease to contribute to the ``running'' of the coupling. 
Note, however, a crucial point, which has been the source of considerable 
confusion in the recent literature:
the ``decoupling'', as described above,   is {\it not} a synonym for non-interactive! 
In the electroweak sector, for example, such a ``decoupling'' takes place indeed, since below the 
mass of the $W$ the gauge boson loops do not contribute to the 
running. However, this is by no means the same as saying that the theory  
is non-interactive (for one thing, the $\beta$ decay still takes place.)
Another central point is that when the QCD charge is constant (non-vanishing!) in the infrared 
(and the quark masses are ignored), QCD becomes conformally invariant, 
and the AdS/CFT correspondence becomes applicable~\cite{Brodsky:2003px}.  
In conclusion:
{\it The  gluon mass keeps QCD strongly coupled, and with the ``conformal window'' open!}


\subsection{A case of SDE-lattice synergy:\\ The Kugo-Ojima function.}

The Kugo-Ojima (KO) scenario~\cite{Kugo:1979gm} claims to 
establish a highly non-trivial link  
between confinement and the infrared behavior of the ghost dressing function $F(q^2)$.
In a nutshell, a  sufficient  condition  for  the realization of 
the KO confinement (``quartet'') mechanism
is that a certain correlation function, $u(q^2)$, defined as 
\bea
&&\int\!d^4x\ \mathrm{e}^{-iq\cdot(x-y)}\langle T\big[\left({\cal D}_\mu c\right)_x^m
\left(f^{nrs}A^n_\nu\bar c^s\right)_y\big]\rangle =
\nonumber\\
&& P_{\mu\nu}(q)\delta^{mn}u(q^2),
\label{KO-1}
\eea
should satisfy the condition $u(0)=-1$. 
Given that $u(0)$ is  related to the  infrared behavior of 
$F(q^2)$ through  the identity $F^{-1}(0)=1+u(0)$, 
the  KO confinement scenario  predicts a
divergent ghost dressing function, and vice-versa.
As was already mentioned, however,    
the ghost dressing function is not enhanced on the lattice~\cite{Cucchieri:2007md,Bogolubsky:2007ud}.
In addition, and perhaps not surprisingly, 
the lattice~\cite{Sternbeck:2006rd} finds no evidence of $u(0)=-1$ either: 
$u(q^2)$ saturates in the deep infrared around approximately  $-0.6$.

\begin{figure*}[t]
\begin{center}
\includegraphics[scale=1.90]{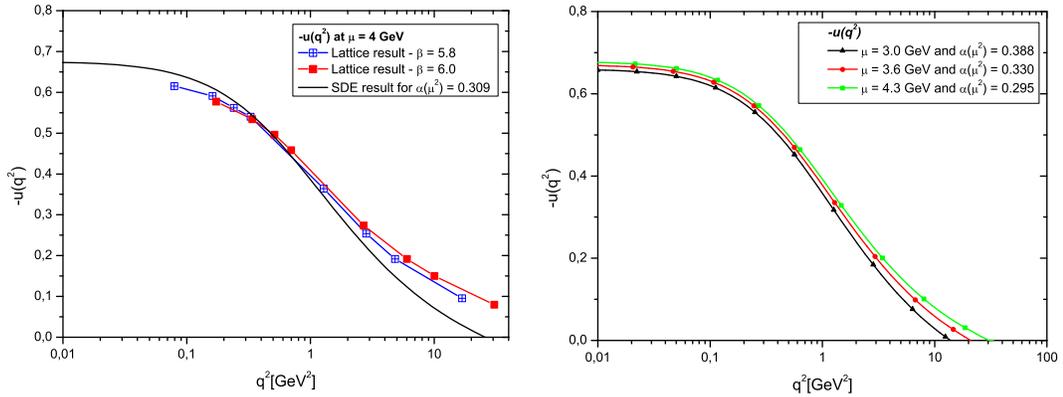}
\vspace{-2.5cm}
\caption{Left: $-u(q^2)$ from SDE-lattice~\cite{Aguilar:2009pp}  
compared to~\cite{Sternbeck:2006rd}.  
Right: The $\mu$-dependence of $-u(q^2)$.}
\label{fig9}
\end{center}
\end{figure*}


Quite remarkably, even though their 
field-theoretic origin appears to be completely different, the KO function 
coincides (in the LG!)  with the $G(q^2)$, introduced in (\ref{LDec}), namely~\cite{Grassi:2004yq} 
\be
u(q^2)=G(q^2). 
\ee
One may then substitute  
into the SDE for  
$G(q^2)$, given in (\ref{sd3}), 
the available lattice data on the gluon and ghost propagators (e.g.~\cite{Bogolubsky:2007ud}), 
and obtain an {\it indirect} determination for $u(q^2)$~\cite{Aguilar:2009pp}. 
Note that from (\ref{sd3}) 
one can verify explicitly~\cite{Aguilar:2009nf} that \mbox{$L(0)=0$}. The 
$u(q^2)$ so obtained may be then compared with the {\it direct}
lattice determination of the same quantity~\cite{Sternbeck:2006rd}, 
from its defining Eq.(\ref{KO-1}). 

The outcome of this procedure is displayed in the left panel of Fig.\ref{fig9}. 
Evidently, the coincidence between the result obtained from the 
combined (indirect) SDE-lattice appraoch and that of the 
pure (direct) lattice is rather good.
Note in addition an important (and only recently appreciated) point: 
$u(q^2)$, and in particular $u(0)$, depend {\it explicitly} 
on the renormalization point $\mu$, 
as one would expect, given that the 
KO function is {\it not} 
an intrinsically  $\mu$-independent quantity [unlike, e.g., $\widehat{d}(q^2)$].
The implications of this, and other facts,  
for the KO confinement mechanism (and approaches relying on it)  
are currently under intense scrutiny~\cite{Dudal:2009xh}.

\section{Conclusions}

In this presentation we have outlined the salient features of the 
SDEs formulated within the PT-BFM framework,  and have 
given several examples of the considerable potential offered  
by their interplay with the lattice simulations.  
Clearly several aspects need to be further investigated, e.g.\\  
{\it (i)}
the dependence of the infrared behavior on the gauge chosen (e.g., LG vs Feynman gauge); \\ 
{\it (ii)} study on the lattice the auxiliary (ghost) Green's functions appearing in the new SDE;\\ 
{\it (iii)} refinement of the SDE treatment to improve the agreement with lattice; \\
{\it (iv)} the possible connection between the SDEs and the Gribov horizon.\\  
We hope that the ongoing effort of the lattice and SDE communities will soon shed light 
on these and many more questions.

{\it Acknowledgments:} 
The author thanks the organizers of LC09 for a most enjoyable workshop. 
This research is supported by the European FEDER and  Spanish MICINN under grant FPA2008-02878, 
and the ``Fundaci\'on General'' of the University of Valencia.

\end{document}